\documentclass[runningheads]{llncs}
\usepackage{graphicx}
\usepackage{amssymb}
\usepackage{rotating} 
\usepackage{xcolor}
\usepackage{amsfonts,amssymb}
\usepackage{algorithm}
\usepackage{algorithmic}
\usepackage{booktabs}
\usepackage{makecell}
\usepackage{bbding}
\usepackage[misc]{ifsym}
\newcommand{\ignore}[1]{}

\begin{document}
\title{IDSGAN: Generative Adversarial Networks for Attack Generation against Intrusion Detection}
\titlerunning{IDSGAN: GAN for Attack Generation against Intrusion Detection}
%
\author{Zilong Lin\inst{1,2}\thanks{This work was done when the author was \ignore{a graduate student }at Shanghai Jiao Tong University.} \and
Yong Shi\inst{1} \and
Zhi Xue\inst{1 (}\Envelope\inst{)}}

\institute{Shanghai Jiao Tong University, Shanghai, China\\
\and
Indiana University Bloomington, Bloomington, IN, USA\\
\email{zillin@indiana.edu, \{shiyong,zxue\}@sjtu.edu.cn}}

\authorrunning{Z. Lin et al.}

%
\maketitle              

\begin{abstract}
As an essential tool in security, the intrusion detection system bears the responsibility of the defense to network attacks performed by malicious traffic. Nowadays, with the help of machine learning algorithms, intrusion detection systems develop rapidly. However, the robustness of this system is questionable when it faces adversarial attacks. For the robustness of detection systems, more potential attack approaches are under research. In this paper, a framework of the generative adversarial networks, called IDSGAN, is proposed to generate the adversarial malicious traffic records aiming to attack intrusion detection systems by deceiving and evading the detection. Given that the internal structure and parameters of the detection system are unknown to attackers, the adversarial attack examples perform the black-box attacks against the detection system. IDSGAN leverages a generator to transform original malicious traffic records into adversarial malicious ones. A discriminator classifies traffic examples and dynamically learns the real-time black-box detection system. More significantly, the restricted modification mechanism is designed for the adversarial generation to preserve original attack functionalities of adversarial traffic records. The effectiveness of the model is indicated by attacking multiple algorithm-based detection models with different attack categories. The robustness is verified by changing the number of the modified features. A comparative experiment with adversarial attack baselines demonstrates the superiority of our model.

\keywords{Generative adversarial networks \and Intrusion detection \and Adversarial examples \and Black-box attack.}

\end{abstract}

\section{Introduction}
\label{intro}
With the spread of security threats on the Internet, the intrusion detection system (IDS) has become the essential tool to detect and defend network attacks in the form of malicious network traffic. The IDS monitors the network by analyzing the features extracted from the network traffic and raises the alarm if unsafe traffic is identified. The main aim of IDS is to audit and classify the network traffic records between normal ones and malicious ones.
As a classification issue, the IDS has widely leveraged machine learning algorithms to classify traffic based on the feature records, including KNN, SVM, Decision Tree, etc~\cite{tsai2009intrusion}. In recent years, deep learning algorithms further contributed with an improvement to IDS in accuracy and simplification~\cite{li2017intrusion}.

However, the classification algorithms expose the vulnerability under the adversarial examples in the recent work, in which the designed adversarial inputs would cause the classifiers' misclassification~\cite{carlini2017adversarial}. For such an attack, the generative adversarial network (GAN)~\cite{goodfellow2014generative} is the potential method for such adversarial example generation. GAN has been implemented in attacks within information security, like malware generation, author attribute anonymity, and password guessing~\cite{hu2017generating,shetty2018a4nt,hitaj2019passgan,pasquini2021improving}.
Although some previous work has applied adversarial learning methods to attack target IDS models~\cite{wang2018deep,yang2018adversarial}, we still hardly know GAN's capability of dynamically attacking multiple IDS models with adversarial malicious traffic with its functionality preserved.

In this paper, we proposed a new framework of GAN, named IDSGAN, for the adversarial attack generation against intrusion detection systems. The goal of the model is to generate malicious feature records of the attack traffic, which can deceive and bypass the detection of the defense systems and, finally, to guide the evasion attack in real networks. Following adversarial malicious traffic records, the attackers can design traffic to evade the detection in real-world attacks. In the model, we designed and improved the generator and the discriminator based on Wasserstein GAN~\cite{arjovsky2017wasserstein}. The generator generates adversarial malicious traffic records. Given that the internal structure and parameters of the black-box IDS are unknown for attackers, the discriminator learns the IDS by its real-time outputs and provides feedback for the training of the generator. We assumed that the outputs of the black-box IDS models can be obtained by querying IDS with traffic records. In the experiment, we modeled multiple IDS models powered by different baseline machine learning algorithms, simulating the intrusion detection systems in reality. In summary, the following contributions are made in this work:
\begin{itemize}
\item An improved framework named IDSGAN is proposed to generate adversarial malicious traffic records to guide the evasion attack against IDS. To preserve the functionality of real traffic generated by IDSGAN, we designed a mechanism that restricts the modification to functional features of original malicious traffic records in adversarial example generation. By dynamically learning the real-time results from IDS models in adversarial training, IDSGAN can attack the updated IDS models powered by different algorithms.
\item We attacked different variants of IDS models to evaluate the effectiveness and robustness of IDSGAN. We measured the attack effectiveness with quantitative evaluations. To verify the robustness, we analyzed the attacks under different numbers of the modified features.
\item We demonstrated that our model outperformed other adversarial attack baselines of intrusion detection.
\end{itemize}

\section{Related Work}
With the rapid development of adversarial learning, adversarial examples generation has attracted researchers' interests and been applied in intrusion detection.

Rigaki leveraged FGSM and JSMA methods to generate the adversarial traffic records, which can evade the detection of IDS, based on the NSL-KDD dataset~\cite{rigaki2017adversarial}. Wang further proposed to apply more adversarial attack algorithms (including JSMA, Targeted FGSM, DeepFool, CW, etc.) to craft adversarial traffic records~\cite{wang2018deep}. Assuming that the attackers have knowledge about the target victim models, both works focused on the white-box adversarial attacks. 

In the research on black-box attacks, Yang proposed zeroth-order optimization and generative adversarial networks to attack IDS~\cite{yang2018adversarial}. However, in this work, the traffic record features were manipulated without the discrimination of features' function, leading to the ineffectiveness of the traffic's attack functionality. Additionally, without learning the latest knowledge of IDS like querying, the discriminator of GAN, a pretrained classifier, cannot dynamically adapt the generation for the attack against the updated IDS models. To solve the above issues, we designed a mechanism to preserve the original attack functionality in different types of malicious traffic, and leveraged the real-time query results from the target IDS to let our model dynamically learn and adapt the target IDS model.
\section{Methodology}
IDSGAN, an improved GAN model, aims to perform the evasion attacks by generating adversarial malicious traffic records, which enable fooling the black-box IDS models. In IDSGAN, the generator and the discriminator engage in a two-player minimax game for adversarial example generation. Based on the characteristics of the benchmark dataset NSL-KDD, it is necessary to preprocess the dataset to fit the model.

\subsection{Dataset: NSL-KDD Dataset Description}
NSL-KDD is used as a benchmark dataset to evaluate IDS today. In NSL-KDD, the dataset comprises the training set KDDTrain+ and the test set KDDTest+. Extracted from the real network environment, the data contains the normal traffic and four main categories of malicious traffic, including Probing (Probe), Denial of Service (DoS), User to Root (U2R), and Root to Local (R2L).

The traffic records in NSL-KDD are extracted into the feature sequences, as the abstract description of the normal and malicious network traffic. Each element in the sequence represents one feature of the traffic. There are 9 features in discrete values and 32 features in continuous values, a total of 41 features. Based on the meanings of each feature, these features consists of four sets including ``\texttt{intrinsic}'', ``\texttt{content}'', ``\texttt{time-based traffic}'', and ``\texttt{host-based traffic}''.

\subsection{Data Preprocessing}
Given multiple feature types and ranges in NSL-KDD, the numeric conversion and the normalization are leveraged for preprocessing data before being fed into models.
Three nonnumeric features (including protocol\_type, service, flag) are embedded. For instance, ``protocol\_type'', including three attributes: TCP, UDP, and ICMP, will be converted into one-hot vectors.
To eliminate the range impact between feature values in the input vectors, the min-max normalization method transforms all original numeric features of the data into the range of $[0,1]$.

\subsection{Structure of IDSGAN}
\begin{figure*}[!t]
\centering
\includegraphics[height=3.94cm]{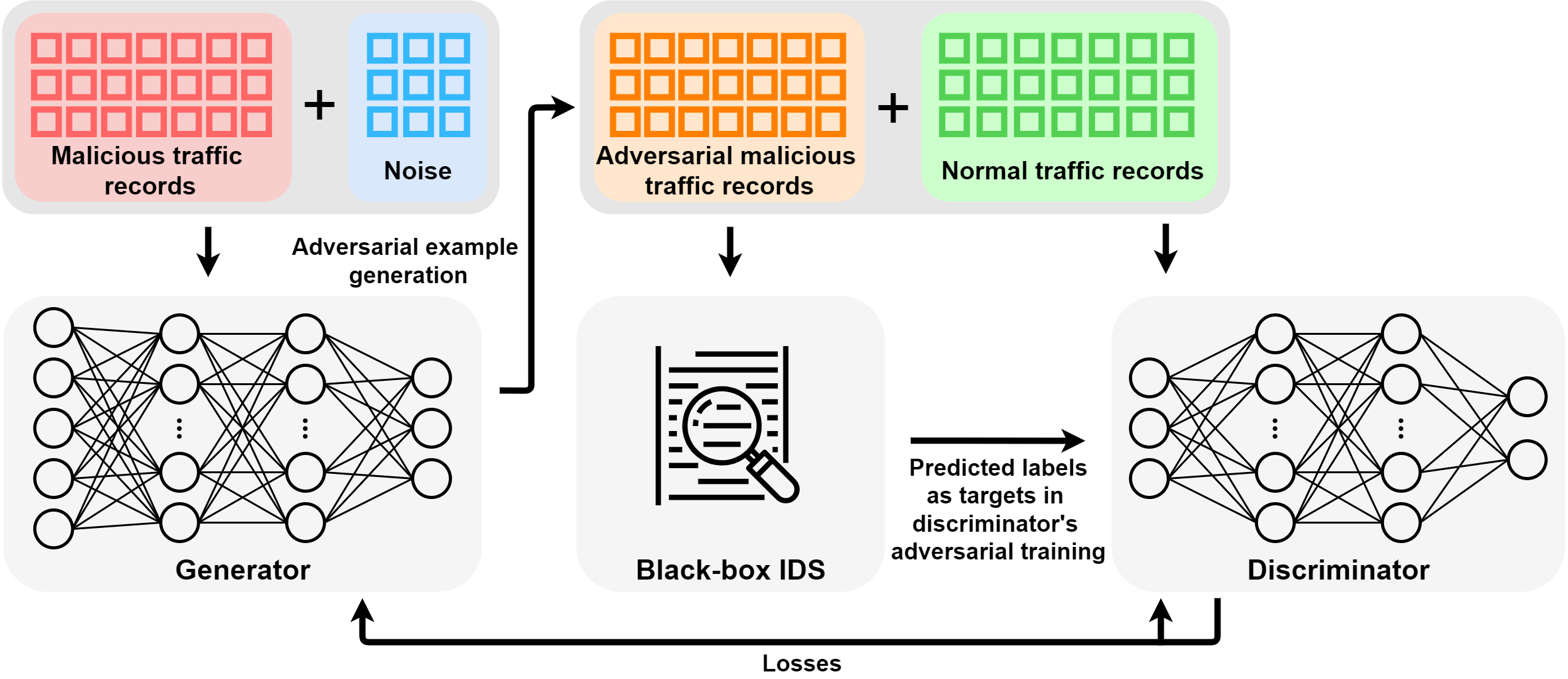}
\caption{The training of IDSGAN. The training set is divided into normal traffic records and malicious traffic records. After adding noise, the malicious records are sent into the generator and modified as the adversarial ones. The black-box IDS predicts the adversarial malicious records and normal ones. The predicted labels are used in the discriminator to learn the black-box IDS. The losses of the generator and discriminator are calculated based on the results of the discriminator and the predicted labels of IDS.
}
\label{fig:IDSGAN_structure}
\vspace{-10pt}
\end{figure*}

Many GAN structures have been designed for different requests. To prevent the non-convergence and instability of GAN, IDSGAN is proposed based on Wasserstein GAN~\cite{arjovsky2017wasserstein}. For the evasion attack against IDS, the generator modifies features to generate adversarial malicious traffic records. Complying with the proposed mechanism of restricted modification, the generation preserves original malicious functionalities of adversarial traffic. The discriminator is trained to learn the black-box IDS and feedbacks the training of the generator. Different machine learning algorithms power the black-box IDS to simulate the IDS models in the real world. The framework of IDSGAN is delineated in Figure \ref{fig:IDSGAN_structure}.

\subsubsection{Restricted modification mechanism}
Although the aim of IDSGAN's generating adversarial attack examples is to evade IDS, the premise is that this generation should retain the original attack functionality of malicious traffic so that such traffic generated based on the adversarial records from IDSGAN can be reproduced and launch network attacks in reality.

Based on the attack principles and purposes, it is evident that each category of attacks has its specific functional features representing the basic functionality of this attack. It means that, in adversarial example generations, the attack attribute would remain unaltered if we solely fine-tuned nonfunctional features, not functional features. Thus, in our mechanism, the functional features of each attack should be kept unchanged to preserve malicious functionalities. The mechanism allows the fine-tuning or retention of nonfunctional features that do not represent the functionality relevant to that attack. These retained features, including functional features, are named the unmodified features in our work. The functional features of each attack category in NSL-KDD are in Table~\ref{table:function_features}.

\begin{table}[!t]
\centering
\caption{The functional features of each attack category.}
\label{table:function_features}  
\begin{tabular}{l|c|c|c|c}
\hline
Attack & Intrinsic & Content & Time-based traffic & Host-based traffic \\
\hline
\hline
Probe & \checkmark &  & \checkmark & \checkmark \\
\hline
DoS & \checkmark &  & \checkmark & \\
\hline
U2R & \checkmark & \checkmark & & \\
\hline
R2L & \checkmark & \checkmark & & \\
\hline
\end{tabular}
\vspace{-10pt}
\end{table}

\subsubsection{Generator}
As a crucial part of the model, the generator plays the role of generating adversarial malicious traffic records for the evasion attack to IDS.

Aiming to transform an original example into an adversarial one, the initial noise perturbation is added to the original traffic record example before the generation. We concatenated the $m$-dimensional original example vector $M$ and the $n$-dimensional noise vector $N$ as an input vector fed into the generator. As the original example part, $M$ has been preprocessed. To be consistent with the normalized vector $M$, the elements of the noise part are randomized in a uniform distribution within the range of $[0,1]$.

Our proposed structure of the generator has a neural network structure with five linear layers. The ReLU non-linearity $F=\max(0,x)$ is utilized to activate the outputs of former four linear layers. To ensure that the adversarial examples share the same dimension as the original example vector $M$, the output layer has $m$ units. The loss of the generator is calculated based on the classification results from the discriminator (see Equation~\ref{equ:loss_G}).

In addition, some tricks exist in the processing of the modified features. To restrict the output elements into the range of $[0,1]$,  we set the element above 1 as 1 and the element below 0 as 0. Concerning that ``\texttt{intrinsic}'' features are the functional features in all the attacks of NSL-KDD, the nonnumeric features will not be modified. For the \ignore{six }binary features in the feature vector, the values of those modified binary features will be transformed into binary values with 0.5 as the threshold after the generator's processing. We transformed the values above the threshold into 1 and those below the threshold into 0.

\subsubsection{Discriminator}
Without the knowledge of the structure and parameters in black-box IDS models, we assumed that the real-time classification results of the black-box IDS models can be obtained by querying. The discriminator is a multi-layer neural network to classify malicious records and normal ones. Also, the discriminator is responsible for learning and imitating the black-box IDS based on the detected samples and their latest predictions from the target IDS. In adversarial training, the normal traffic records and the adversarial malicious traffic records are first classified by the black-box IDS. Then, for the imitation to the black-box IDS, the same dataset labeled by the target IDS is shared to the discriminator as the training set whose current labels are the real-time predictions from this IDS, shown in Figure~\ref{fig:IDSGAN_structure}. Reflecting the dynamic optimization of the structure and parameters in the black-box IDS, the IDS's real-time predictions are leveraged for the discriminator to learn the IDS dynamically.

Additionally, the discriminator helps the training of the generator, whose gradient is back-propagated from the discriminator. Based on the loss calculated by the real-time results of the discriminator and the black-box IDS, the generator dynamically optimizes the evasion strategy to fine-tune malicious records, enabling adversarial examples from IDSGAN to attack the real-time IDS models.

\begin{algorithm}[!t]
\scriptsize
\caption{IDSGAN}
\label{alg:IDSGAN}
\begin{algorithmic}[1] 
\REQUIRE ~~\\ 
Original normal and malicious traffic records $S_{normal}$, $S_{attack}$;\\
The noise $N$ for the adversarial generation;\\
The pretrained black-box IDS $B$;\\
\ENSURE ~~\\
The optimization of the generator $G$ and the discriminator $D$;\\
\STATE Initialize the generator $G$ and the discriminator $D$;
\label{ code:initialize }
\FOR{number of training iterations}
    \FOR{$G$-steps}
        \STATE $G$ generates the adversarial malicious traffic examples based on $S_{attack}$;
        \STATE Update the parameters of $G$ according to Eq. \ref{equ:loss_G};
    \ENDFOR
    \label{code:G-step}
    \FOR{$D$-steps}
        \STATE $B$ classifies the training set ($S_{normal}$, $G(S_{attack},N)$), getting predicted labels;
        \STATE $D$ classifies the same training set ($S_{normal}$, $G(S_{attack},N)$);
        \STATE Update the parameters of $D$ according to Eq. \ref{equ:loss_D};
    \ENDFOR
    \label{code:D-step}
\ENDFOR
\end{algorithmic}
\end{algorithm}

\subsubsection{Training algorithms}
In the generator's training, the discriminator's detection results to the adversarial examples provide the gradient information for the generator. The loss function of the generator is defined in Equation \ref{equ:loss_G}.
\begin{equation}
L_{G}=\mathbb{E}_{M\in{S_{attack},N}}D(G(M,N))
\label{equ:loss_G}
\end{equation}
where $S_{attack}$ is the original malicious traffic records; $G$ and $D$ represent the generator and the discriminator, respectively. To train and optimize the generator for fooling the black-box IDS, we need to minimize $L_{G}$.

The training set of the discriminator includes adversarial malicious records and normal records. With the aim of learning the black-box IDS, the discriminator gets the loss calculated by the output labels of the discriminator and the predicted labels achieved from the black-box IDS. Thus, the loss function for the discriminator's optimization is in Equation \ref{equ:loss_D}.
\begin{equation}
L_{D}=\mathbb{E}_{s\in{B_{normal}}}D(s)-\mathbb{E}_{s\in{B_{attack}}}D(s)
\label{equ:loss_D}
\end{equation}
where $s$ means the training set of the discriminator; $B_{normal}$ and $B_{attack}$ mean the normal traffic records and the adversarial traffic records with the predicted labels from the black-box IDS as the ground truth, respectively.

Based on Wasserstein GAN, RMSProp is the optimizer of IDSGAN to optimize the parameters in the model. The outline of the IDSGAN adversarial training is shown in Algorithm \ref{alg:IDSGAN}.

\ignore{
\begin{algorithm}[!t]
\caption{IDSGAN}
\label{alg:IDSGAN}
\begin{algorithmic}[1] 
\REQUIRE ~~\\ 
Original normal and malicious traffic records $S_{normal}$, $S_{attack}$;\\
The noise $N$ for the adversarial generation;\\
\ENSURE ~~\\
The optimization of the generator $G$ and the discriminator $D$;\\
\STATE Initialize the generator $G$, the discriminator $D$ and the black-box IDS $B$;
\label{ code:initialize }
\FOR{number of training iterations}
    \FOR{$G$-steps}
        \STATE $G$ generates the adversarial malicious traffic examples based on $S_{attack}$;
        \STATE Update the parameters of $G$ according to Eq. \ref{equ:loss_G};
    \ENDFOR
    \label{code:G-step}
    \FOR{$D$-steps}
        \STATE $B$ classifies the training set including $S_{normal}$ and $G(S_{attack},N)$, getting predicted labels;
        \STATE $D$ classifies the training set;
        \STATE Update the parameters of $D$ according to Eq. \ref{equ:loss_D};
    \ENDFOR
    \label{code:D-step}
\ENDFOR
\end{algorithmic}
\end{algorithm}

}
\section{Empirical Evaluation}

\subsection{Experimental Setup}
\label{experimental setup}
In our implementation, we utilized \texttt{PyTorch} as the deep learning framework to construct the IDSGAN model\ignore{~\cite{paszke2017pytorch}}. We built black-box IDS models by \texttt{scikit-learn}. The proposed model was evaluated on a Linux PC with Intel Core i7-2600.

IDSGAN is trained with a 64 batch size for 100 epochs. The learning rates of the generator and discriminator are both 0.0001. The weight clipping threshold in the discriminator's training is 0.01. The dimension of the noise vector is 9. 

Based on the relevant researches in intrusion detection, multiple machine learning algorithms have been used in IDS. To evaluate the capacity and generalization of our purposed model against IDS comprehensively, we built seven types of algorithm-based black-box IDS models which are commonly applied as the baseline approaches to validate improved intrusion detection systems \cite{tsai2009intrusion}. The algorithms of the black-box IDS models adopted in the evaluation include Support Vector Machine (SVM), Naive Bayes (NB), Multilayer Perceptrons (MLP), Logistic Regression (LR), Decision Tree (DT), Random Forest (RF), and K-Nearest Neighbors (KNN). In the evaluation, the black-box IDS models have been pretrained with their training set before generating adversarial samples.

The training and test sets are designed based on the NSL-KDD dataset (containing KDDTrain+ and KDDTest+). The training set for the black-box IDS consists of one-half of the records in KDDTrain+, containing normal and malicious traffic records. The training set of the discriminator includes the normal traffic records in the other half of KDDTrain+ and the adversarial malicious records from the generator. Given that the nonfunctional features to be modified in each attack category are different, IDSGAN generates the adversarial examples for solely one attack category each time. So, the training set of the generator for one attack category is the records of that category in the other half of KDDTrain+. The records of one attack category in KDDTest+ are the test set for the generator, which aims for the adversarial samples of that category.

For the experimental metrics, the detection rate and the evasion increase rate are calculated, presenting the performance of IDSGAN directly and comparatively. The detection rate (i.e., DR) is the proportion of correctly detected malicious traffic records by the black-box IDS to all of those attack records, directly showing the evasion ability of the adversarial examples and the robustness of the black-box IDS. The original and adversarial detection rates represent the detection rate of the original malicious traffic records and that of the adversarial malicious traffic records, respectively. In addition, we defined the evasion increase rate (i.e., EIR) as the rate of the increase in the undetected malicious examples by IDS between original malicious examples and adversarial examples, evaluating the evasion attack of IDSGAN. These metrics are calculated as follows.

\begin{equation}
DR=\frac{Num.\ of\ correctly\ detected\ attacks}{Num.\ of\ all\ the\ attacks}
\label{equ:DR}
\end{equation}

\begin{equation}
EIR=1-\frac{Adversarial\ detection\ rate}{Original\ detection\ rate}
\label{equ:EIR}
\end{equation}

A lower detection rate means more malicious traffic records evade the detection of the black-box IDS, revealing a higher adversarial attack effectiveness. On the contrary, a lower evasion increase rate reflects a low increase rate on the number of adversarial examples evading the black-box IDS, meaning a low improvement in the evasion attack capacity of adversarial examples compared with the original ones. So, the goal for the IDSGAN optimization is to obtain a lower detection rate and a higher evasion increase rate.

\subsection{Effectiveness in Different Attack Categories}
\label{sect:Measurement_in_diff_attacks}
\begin{table}[!t]
\centering
\caption{The performance of IDSGAN under DoS, U2R and R2L. The first two lines are the black-box IDS's original detection rates to the original test set. In Column ``Add'', ``$\times$'' means the adversarial generation with only the functional features unmodified, while ``\checkmark'' means that the unmodified features are added in the generation.} 
\label{table:measurement}  
\begin{tabular}{c|l|l|c|c|c|c|c|c|c}
\hline
Add & Attack & Metric (\%) & SVM & NB & MLP & LR & DT & RF & KNN \\
\hline
\hline
-- &	DoS & Original DR & 82.37 & 84.94 & 82.70 & 79.85 & 75.13 & 73.28 & 77.22\\
\hline
-- & U2R \& R2L & Original DR & 0.68 & 6.19 & 4.54 & 0.64 & 12.66 & 2.44 & 5.69\\
\hline
$\times$ & DoS & Adversarial DR & 0.46 & 0.01 & 0.72 & 0.36 & 0.20 & 0.35 & 0.37\\
\hline
$\times$ & DoS & EIR & 99.44 & 99.99 & 99.13 & 99.55 & 99.73 & 99.52 & 99.52\\
\hline
$\times$ & U2R \& R2L & Adversarial DR & 0.00 & 0.01 & 0.00 & 0.00 & 0.02 & 0.00 & 0.00\\
\hline
$\times$ & U2R \& R2L & EIR & 100.00 & 99.84 & 100.00 & 100.00 & 99.84 & 100.00 & 100.00\\
\hline
\checkmark & DoS & Adversarial DR & 1.03 & 1.26 & 1.21 & 0.97 & 0.36 & 0.77 & 1.16\\
\hline
\checkmark & DoS & EIR & 98.75 & 98.52 & 98.54 & 98.79 & 99.52 & 98.95 & 98.50\\
\hline
\checkmark & U2R \& R2L & Adversarial DR & 0.01 & 0.08 & 0.01 & 0.00 & 0.07 & 0.00 & 0.00\\
\hline
\checkmark & U2R \& R2L & EIR & 98.53 & 98.71 & 99.78 & 100.00 & 99.45 & 100.00 & 100.00\\
\hline
\end{tabular}
\vspace{-10pt}
\end{table}

To evaluate the model comprehensively, the trained IDSGAN was tested to generate the adversarial malicious samples based on KDDTest+. Given that DoS and Probe are both attacks based on network, we only tested on DoS to show the performance of IDSGAN on such kinds of attacks. Also, the attacks based on the traffic content like U2R and R2L were tested. Due to the similar characteristics shared among U2R and R2L---leading to the same functional features among two categories---and their small data amount in the dataset, we gathered U2R and R2L as one attack group in our work.

Before the generation by IDSGAN, the original detection rates to DoS, U2R, and R2L were calculated on the trained black-box IDS, shown in Table \ref{table:measurement}. By reason of the small number of U2R and R2L records in the training set, the insufficient learning makes the low original detection rates to U2R and R2L. A similar performance has been reported in the previous work~\cite{sapre2019robust,ingre2017decision}.

\begin{figure}[!t]
    \centering
        \begin{minipage}[t]{0.45\textwidth}
            \centering
            \includegraphics[height=3.7cm]{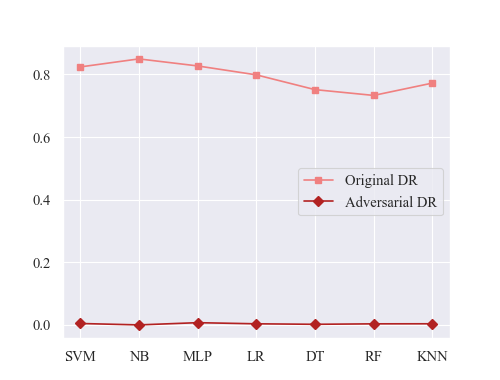}\\ 
            (a)
        \end{minipage}
        \begin{minipage}[t]{0.45\textwidth}
            \centering
            \includegraphics[height=3.7cm]{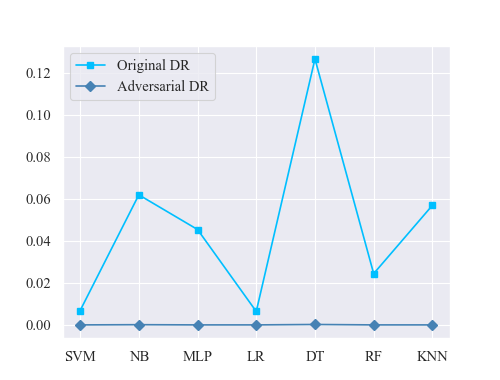}\\
            (b)
        \end{minipage}
    \caption{The comparisons of the adversarial detection rates and the original detection rates under different black-box IDS models with only the functional features unmodified. (a) is the results of DoS and (b) is the results of U2R and R2L.}
    \label{fig:figure4}
\vspace{-10pt}
\end{figure}

First, to evaluate the effectiveness of IDSGAN, we tested the capacity of IDSGAN in different attacks with only the functional features unmodified. In the experiment results shown in Table \ref{table:measurement} and Figure \ref{fig:figure4}, all of the adversarial detection rates to DoS, U2R, and R2L under different IDS algorithms decline compared with the original detection rates and reach around 0, indicating that the IDS models are almost incapable of classifying any adversarial examples.

As shown in Figure \ref{fig:figure4}(a), the adversarial detection rates to DoS under all detection algorithms remarkably decrease from around 80\% to less than 1\%. Although Multilayer Perceptrons shows the best robustness in the list of all IDS models, its adversarial detection rate to DoS is only 0.72\%. More than 99.0\% of the adversarial DoS traffic examples evade the detection of the black-box IDS model in each test. The evasion increase rate in each DoS test is above 99.0\%. The results indicate the excellent effectiveness of IDSGAN in DoS.

For U2R and R2L in Figure \ref{fig:figure4}(b), while the difference of the original detection rates between algorithms is noticeable, all of the adversarial detection rates are equal to or close to 0, indicating almost all of the original detectable examples of U2R and R2L can fool and evade the IDS after the adversarial generation. With a significant increase in the adversarial malicious examples fooling the IDS, the evasion increase rates are also high, all of which are above 99.5\%.

The low adversarial detection rates and high evasion increase rates obtained under various attack categories and multiple IDS algorithms indicate IDSGAN's great effectiveness and generalization in adversarial attacks of evading various IDS models. Besides, some tiny differences still exist in the performance under different attack categories and different IDS models.

\begin{figure}[!t]
    \centering
        \begin{minipage}[t]{0.45\textwidth} 
            \centering
            \label{fig:figure5a}
            \includegraphics[height=3.7cm]{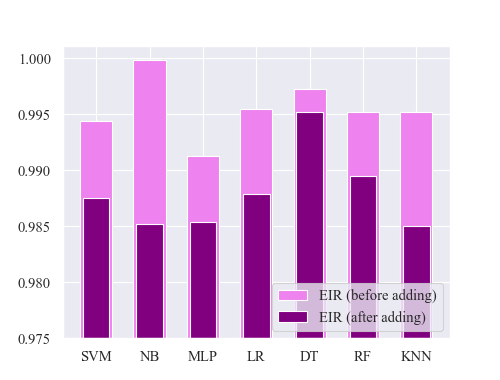}\\
            (a)
        \end{minipage}
        \begin{minipage}[t]{0.45\textwidth} 
            \centering
            \label{fig:figure5b}
            \includegraphics[height=3.7cm]{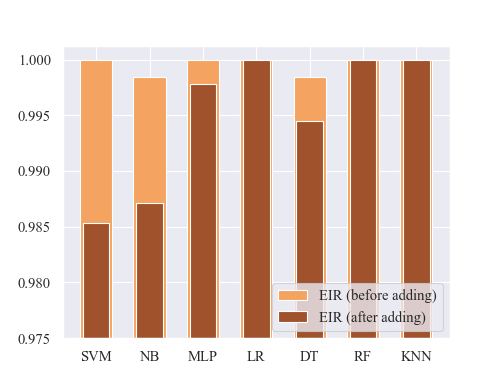}\\
            (b)
        \end{minipage}
    \caption{The comparisons of the evasion increase rates under various algorithms of black-box IDS models before adding unmodified features and after adding unmodified features. (a) is the results of DoS and (b) is the results of U2R and R2L.}
    \label{fig:figure5}
\vspace{-10pt}
\end{figure}

\subsection{Robustness with Different Numbers of Modified Features}
The number of the modified features is an influential factor that significantly affects our proposed model's success in adversarial attacks. To evaluate its robustness, we conducted the contrast tests on DoS, U2R, and R2L, altering the number of the modified features and measuring such impact. Given that the attack with the fewest unmodified features is the scenario with only retaining this attack's functional features representing the attack function, the way to alter the number of modified features is adding some nonfunctional features into the group of the unmodified features. All of the nonfunctional features were modified features in Section \ref{sect:Measurement_in_diff_attacks}. We randomly picked 50\% of features from each nonfunctional feature set as the added unmodified features in the contrast tests.

In Table \ref{table:measurement} and Figure \ref{fig:figure5}, the evasion increase rates decrease slightly or maintain in contrast experiments, compared with the results in Section \ref{sect:Measurement_in_diff_attacks} with only the functional features unmodified. In Figure \ref{fig:figure5}, the minor changes in metric results indicate IDSGAN's strong robustness under fewer modified features. The strong perturbation to modified features still causes high evasion increase rates for all IDS algorithms, even if lowering the number of modified features. For instance, the evasion increase rates keep unchanged after adding unmodified features when Logistic Regression detects U2R and R2L.

However, with the increase of unmodified features, more original information of a traffic record is retained in the adversarial generation, leading to the rise of accuracy in IDS detection. Besides, the improvements in detection results are different under various attack categories and IDS algorithms.
Considering the detection improvement's relatedness with attack categories, the evasion increase rate of DoS under K-Nearest Neighbors decreases about 1.00\% after changing the modified features in Figure \ref{fig:figure5}(a). However, no variation occurs in results when that algorithm detects U2R and R2L in Figure \ref{fig:figure5}(b). 
Also, for the relatedness with IDS algorithms, the evasion increase rates under Naive Bayes in two attacks change more remarkably than others, revealing the smaller number of modified features has a more significant impact on the evasion against Naive Bayes.

Thus, although the number reduction of the modified features slightly causes more failure in adversarial malicious records bypassing the IDS detection, the sustained high evasion increase rates verify the strong robustness of IDSGAN in the evasion attack. Also, the result shows that the robustness performance of adversarial evasion attacks relies on attack categories and IDS algorithms.

\subsection{Baseline Comparisons}
Besides GAN, different adversarial learning methods have been proposed and applied to help traffic records evade the detection of IDS~\cite{wang2018deep,rigaki2017adversarial,pacheco2021adversarial}. We compared our proposed approach with the following competitive baseline attack models: JSMA Attack, Targeted FGSM Attack, DeepFool Attack, and CW Attack. 

For a fair comparison of the baselines, the intrusion detection system in all the experiments is Multilayer Perceptrons. The detection system and attack models share the same architecture and hyper-parameters as the setting in the previous work~\cite{wang2018deep}, while the proposed restricted modification mechanism takes action to keep traffic functionality. Leveraging the dataset described in Section~\ref{experimental setup}, we trained and tested the models in the experiments. Table~\ref{table:baseline} summarizes the malicious traffic record detection results of adversarial attacks. Our proposed model outperforms all the baselines by a wide margin, and the attack results are notable in both malicious traffic categories. In addition, the results have observable differences between various adversarial attack models, in which JSMA Attack and CW Attack are less devastating than other attack approaches.

We also compared with the GAN baselines in~\cite{yang2018adversarial}, which statically attacked the target IDS without the restriction on the feature modification. The restricted modification mechanism leads the GAN's capacity of evading IDS in~\cite{yang2018adversarial} to drop quickly. The detection rates rise from 24.34\% to 32.45\% in DoS and from 2.23\% to 3.39\% in U2R and R2L. Also, IDSGAN's dynamical imitation strategy by querying the target IDS strengthens evasion attacks remarkably. Compared with~\cite{yang2018adversarial}, IDSGAN outperforms in evasion effectiveness and attack functionality preservation significantly.

\begin{table}[!t]
\scriptsize
\centering
\caption{Detection rates on different adversarial attack approaches}
\label{table:baseline}  
\begin{tabular}{c|c|c|c|c|c|c|c|c}
\hline
Attack & Original & JSMA & FGSM & DeepFool & CW & \makecell{Unrestricted\\ static GAN} & Static GAN & IDSGAN \\ 
\hline
\hline
DoS & 79.12\% & 20.75\% & 7.19\% & 15.86\% & 24.67\% & 24.34\% & 32.45\% & 0.61\%\\
\hline
U2R \& R2L & 4.78\% & 3.06\% & 0.15\% & 1.04\% & 3.20\% & 2.23\% & 3.39\% & 0.00\%\\
\hline
\end{tabular}
\vspace{-10pt}
\end{table}

\section{Conclusion and Future Work}
IDSGAN is a novel framework of generative adversarial networks aiming to generate adversarial attacks that can evade IDS. The model design and the restricted modification mechanism enable IDSGAN to attack against real-time black-box IDS models powered by multiple machine learning algorithms and preserve traffic's malicious functionalities, respectively. In the evaluation, IDSGAN shows its effectiveness in generating adversarial malicious traffic records of different attacks, lowering the detection rates of various IDS models to around 0\%. In the robustness evaluation, the evasion capacity of adversarial malicious examples maintains or slightly reduces after limiting the number of modified features, indicating the model's strong robustness. Also, the comparisons with other adversarial attack methods demonstrate its better performance.

We focused on generating the adversarial malicious traffic records capable of evading the target IDS in this work. In the next step, depending on such adversarial examples, we would produce the malicious network traffic, whose features match the adversarial traffic records, to experimentally attack the running IDS after being approved by our institutional Ethics Board.

%
%
%
\bibliographystyle{splncs04}
\bibliography{refs}

\end{document}